\documentclass[letterpaper,twocolumn,10pt]{article}
\usepackage{usenix2019_v3}

\hypersetup{
     colorlinks = false,
     linkcolor = blue,
     anchorcolor = blue,
     citecolor = blue,
     filecolor = blue,
     urlcolor = blue
     }

\usepackage{graphicx}
\usepackage{tikz}
\usepackage{amsmath}

\usepackage{color,colortbl}
\definecolor{Gray}{gray}{0.9}
\usepackage{paralist}
\usepackage{url}
\usepackage{csquotes}
\usepackage{booktabs,tabularx}
\newcolumntype{Y}{>{\centering\arraybackslash}X}
\renewcommand{\paragraph}[1]{\textbf{#1}}

\usepackage{fontawesome}

\usepackage{adjustbox}

\usepackage{todonotes}

\usepackage[sort,numbers]{natbib}

\usepackage{mwe}

\usetikzlibrary{positioning,shapes,arrows,calc,fit}
\definecolor{rubblue}{RGB}{0,53,96}
\definecolor{rubred}{RGB}{230,51,42}
\definecolor{rubgreen}{RGB}{141,174,16}
\usepackage{standalone}

\usepackage{glossaries}

\newglossaryentry{alexa1m}{name={Alexa top-1 million},description={}}
\newglossaryentry{metadata-attacker}{name={Metadata-Attacker},description={}}

\newacronym{dom}{DOM}{Document Object Model}
\newacronym{xss}{XSS}{Cross-Site Scripting}
\newglossaryentry{sql}{name=SQL,description={}}
\newacronym{sqli}{SQLi}{SQL injection}
\newacronym{cms}{CMS}{Content Management System}
\newacronym{vel}{VEL}{Vulnerable Extensions List}
\newacronym{csp}{CSP}{Content Security Policy}
\newacronym{mvc}{MVC}{Model-View-Controller}
\newacronym{xml}{XML}{eXtensible Markup Language}
\newacronym{domxss}{DOMXSS}{DOM-based XSS}
\newacronym{owasp}{OWASP}{Open Web Application Security Project}
\newacronym{iptc}{IPTC}{International Press Telecommunications Council}
\newacronym{hpkp}{HPKP}{HTTP Public Key Pinning}
\newacronym{hsts}{HSTS}{Hypertext Strict Transport Security}
\newacronym{sri}{SRI}{Sub-resource Integrity}
\newacronym{bamc}{BAMC}{Block-All-Mixed-Content}
\newacronym{uir}{UIR}{Upgrade-Insecure-Requests}
\newacronym{gc}{GC}{Google Chrome}
\newacronym{ie}{IE}{Internet Explorer}

\newglossaryentry{html}{name=HTML, description=Hypertext Markup Language}
\newglossaryentry{joomla}{name=Joomla, description=Content Mangement System}
\newglossaryentry{drupal}{name=Drupal, description=Content Mangement System}
\newglossaryentry{wp}{name=WordPress, description=Content Mangement System}
\newglossaryentry{js}{name=JavaScript, description=JavaScript}
\newglossaryentry{useragent}{name=useragent, description=Useragent}
\newglossaryentry{wordpress}{name=Wordpress, description=Wordpress}
\newglossaryentry{csrf}{name=CSRF, description=}
\newglossaryentry{ajax}{name=AJAX, description=}

\usepackage{listings}

\usepackage{xcolor}

\definecolor{codegreen}{rgb}{0,0.6,0}
\definecolor{codegray}{rgb}{0.5,0.5,0.5}
\definecolor{codepurple}{rgb}{0.58,0,0.82}
\definecolor{backcolour}{rgb}{0.95,0.95,0.92}

\lstdefinestyle{mystyle}{
  backgroundcolor=\color{backcolour},   commentstyle=\color{codegreen},
  keywordstyle=\color{magenta},
  numberstyle=\tiny\color{codegray},
  stringstyle=\color{codepurple},
  basicstyle=\ttfamily,
  breakatwhitespace=false,         
  breaklines=true,                 
  captionpos=b,                    
  keepspaces=true,                 
  numbers=left,                    
  numbersep=5pt,                  
  showspaces=false,                
  showstringspaces=false,
  showtabs=false,                  
  tabsize=2
}

\begin{document}

\date{}

\title{Over 100 Bugs in a Row:\linebreak
  Security Analysis of the Top-Rated \glsentrytext{joomla} Extensions}

\author{
{\rm Marcus Niemietz, Mario Korth, Christian Mainka, Juraj Somorovsky}\\
firstname.lastname@hackmanit.de\\
Hackmanit GmbH
} 

\maketitle

\begin{abstract}

Nearly every second website is using a \gls{cms} such as \gls{wp}, \gls{drupal}, and \gls{joomla}. These systems help to create and modify digital data, typically within a collaborative environment. One common feature is to enrich their functionality by using extensions. Popular extensions allow developers to easily include payment gateways, backup tools, and social media components.

Due to the extended functionality, it is not surprising that such an expansion of complexity implies a bigger attack surface. In contrast to \gls{cms} core systems, extensions are usually not considered during public security audits. However, a \gls{xss} or \gls{sqli} attack within an activated extension has the same effect on the security of a \gls{cms} as the same issue within the core itself. Therefore, vulnerabilities within extensions are a very attractive tool for malicious parties.

We study the security of \gls{cms} extensions using the example \gls{joomla}; one of the most popular systems. We discovered that nearly every second installation of such a system also includes \gls{joomla}'s official top-10 rated extensions as a per se requirement. Moreover, we have detected that every single extension of the official top-10 rated extensions is vulnerable to \gls{xss} and 30\% of them against \gls{sqli}. We show that our findings are not only relevant to \gls{joomla}; two of the analyzed extensions are available within systems like \gls{wp} or \gls{drupal}, and introduce the same vulnerabilities. Finally, we pinpoint mitigation strategies that can be realized within extensions to achieve the same security level as the core \gls{cms}.
\end{abstract}

\section{Introduction}

More than half of the \gls{alexa1m} websites are using a \gls{cms} \cite{w3techs}. Among these systems, \gls{wp}, \gls{joomla}, and \gls{drupal} are the most popular brands with a combined \gls{cms} market share of over 69\%. It is therefore an important task to analyze their security.

These \gls{cms}s are frequently audited by third parties due to their importance, especially widely deployed software like \gls{wp} and \gls{joomla}. There exists a family of tools which, inter alia, allow researchers and penetration testers to scan \gls{cms} software (e.g., WPScan~\cite{WPScan}, JoomScan~\cite{joomscan}) for well-known issues. Although community-based projects like \gls{joomla} make use of tools like static source code scanners, it seems to be a non-trivial task to detect well-known vulnerabilities such as \gls{xss} and \gls{sqli}, as shown by security researches multiple times \cite{Trunde:2015:WSA:2837185.2837195}. To provide an example, \gls{joomla} introduced the usage of the code analysis tool RIPS in June 2018 \cite{184419, joomlaNewsRips}. Although RIPS is explicitly designed to detect web vulnerabilities like \gls{xss}, \gls{joomla} has officially announced 30 security vulnerabilities within the first half year of 2019, and 12 of them were \gls{xss} issues \cite{JomlaSecurityAnnouncements}.

\paragraph{On the Importance of \gls{cms} Extensions.}
Nearly every \gls{cms} includes multiple extensions which add new functionality to the core systems. However, the \gls{cms} extensions also enlarges the attack surface.
As an important aspect, the core of the \gls{cms} and the extension share the same privileges.
Due to the missing isolation, a vulnerability in a popular extension might have the same impact as a vulnerability within the \gls{cms} itself. 

Despite their security importance, \gls{cms} extensions are, in contrast to the \gls{cms} core systems, usually not covered by public security audits.
To highlight this gap, we analyzed the security of extensions from one of the most popular systems called \gls{joomla}. 

\paragraph{Our Approach.}
Due to the high number of nearly 8,500 \gls{joomla} extensions, we restricted our focus to the top-10 rated and freely available extensions within \gls{joomla}'s official extension directory. Moreover, we concentrated on the two top-rated \gls{owasp} injection risks namely \gls{xss} and \gls{sqli}. Our aim is to answer the following research questions:

\begin{enumerate}
    \item How many \gls{joomla} installations have deployed the top-10 rated extensions among the \gls{alexa1m} websites?
	\item Are \gls{joomla} extensions vulnerable to the well-known attacks \gls{xss} and \gls{sqli}? If so, what is the root cause of such security issues and what can we learn from them?
    \item Can we assume that at least one top-rated and maybe vulnerable extension is always activated within \gls{joomla} installations? If so, does a vulnerability within an extension have the same importance as a vulnerability within the \gls{cms} core?
\end{enumerate}

If extensions are a tool which is always available and activated within a \gls{cms}, this would imply that we should always consider extensions with the same priority as the \gls{cms} core. Otherwise, an extension might be the weakest link of a complex chain if it is vulnerable against attacks.

\paragraph{Empirical Study.}
We detected that all analyzed extensions are vulnerable to \gls{xss} and 30\% of them are vulnerable to \gls{sqli}. We responsibly reported and disclosed 103 vulnerabilities across multiple \gls{cms}'s whereas 86 vulnerabilities were within \gls{joomla} extensions. By investigating extensions which share the same code and, therefore, the same vulnerabilities for multiple \glspl{cms}, we identified 11 vulnerabilities within \gls{wp} extensions and 6 vulnerabilities within \gls{drupal} extensions.

\paragraph{Contributions.}
Overall, we make the following contributions:
\begin{itemize}
    \item We detected that \textit{all} of the top-10 rated \gls{joomla} extensions are vulnerable to \gls{xss} and 30\% of them are vulnerable to \gls{sqli}.
	\item We show that 40.9\% of the \gls{joomla} installations within the \gls{alexa1m} websites include at least one of our tested top-rated extensions and are, thus, vulnerable to the discovered attacks.
	\item We developed an open-source tool for detecting \gls{xss} vulnerabilities in web applications which use metadata within media files.
	\item We identify possible root causes of the detected vulnerabilities and demonstrate generalizability by evaluating that extensions which are available for multiple \glspl{cms} also have the same security vulnerabilities.
\end{itemize}

\paragraph{Artifact Availability.} In the interest of open science, we have published our tool Metadata-Attacker as open-source online.
\section{Foundations}

\subsection{\glsentrylong{xss}} 
We consider all three \gls{xss} variants that are mentioned as OWASP top-10 injection threats as relevant: reflected \gls{xss}, stored \gls{xss}, and reflective \gls{domxss}~\cite{owaspXSS}.\footnote{We left out persistent \gls{domxss}~\cite{DBLP:conf/ndss/SteffensRJS19}, because one of its requirements is to have another \gls{xss} to write into the persistent client-side storage. This is covered by seeking for reflective \gls{xss} and \gls{domxss}.}

\textit{Reflected \glsentrylong{xss}.} The first step of a reflective \gls{xss} attack is usually to send attack vectors defined by the attacker to the server via an HTTP GET or POST request. After analyzing the server response message, the attacker checks whether the attacker's code was (partially) displayed and therefore reflected. If this is not the case, the attacker can try another testing code; otherwise, the attacker will abort the attack on the given resource.

The successfully executed testing code can show characters that are necessary for the attack, for example if \lstinline|<| or \lstinline|"| are not filtered by the server. In the case of an HTTP GET request, the attacker can then prepare a link pointing to the vulnerable web application with a malicious JavaScript code embedded in the query string.
In the event of an HTTP POST request, a prepared HTML form can be used, which is auto-submitted as soon as the attack page is loaded.

\textit{Stored \glsentrylong{xss}.} As another server-side vulnerability, stored \gls{xss} occurs when  the attack vector can be stored persistently in the vulnerable web application (e.g., malicious code inside a log file or database). Each time a victim visits the part of the web application where the vector is saved, it is executed. 

\textit{\glsentrylong{domxss}.} Contrary to reflected and stored \gls{xss}, the cause for \gls{domxss} lies in the browser of the user itself; for example, client-side JavaScript code provided by the server.
This JavaScript code takes some input of the client and at some point uses it at a sink such as \lstinline|innerHTML|.
This can result in the injection of arbitrary \gls{html} or JavaScript code and therefore \gls{xss}.

\subsection{SQL Injection} An \gls{sqli} occurs when improperly sanitized user input is used in the construction of \gls{sql} queries.
The problem arises usually when the whole query which was constructed by the application is submitted to the \gls{sql} server instead of the query with placeholders for the parameters (e.g., prepared statements~\cite{preparedStatements}).
Therefore, the \gls{sql} server cannot distinguish between the input of a trusted and a malicious user; for example, selecting another database although just an integer was expected. Hence it simply parses the query and executes it, which in the end can lead to an \gls{sqli} if an attacker was able to insert arbitrary statements.

\subsection{\glsentrytext{joomla}} 
Our tested \gls{cms} called  \gls{joomla} uses a strict \gls{mvc} architecture. \gls{joomla}'s extensions can be either a module, component, plugin, template, or language~\cite{extensionTypesJoomla}.
Beyond the applications base directory, there are common folder paths for every \gls{joomla} installation and its extensions.
Additionally, unless explicitly changed or replaced, each \gls{joomla} installation contains a \lstinline|robots.txt| and \lstinline|web.config.txt| file which contain identifying strings.
Some native components are delivered with identifying \gls{xml} files.
Hence it is possible to identify a \gls{joomla} installation by requesting some of those identifying files.
\section{Methodology}

\subsection{Identifying \glsentrytext{joomla}}

For identifying \glsentrytext{joomla} installations and their extensions, we conducted a large-scale analysis of the \gls{alexa1m} to find out how many websites use \gls{joomla} as a \gls{cms}.
In this regard, we wrote a crawler to detect \gls{joomla} installations with a standard core system (cf., \autoref{sec:noinstall}).
With our crawler, we detected 19,047 \gls{joomla} installations in total.

Based on our results, our remaining task was to determine the number of extensions within the set of our detected installations. Since there are over 8,400 \gls{joomla} extensions available, we restricted our research to freely downloadable and thus non-paid extensions.
This criteria fits to 3,981 out of 8,457 extensions. Among the set of freely downloadable extensions, we took the top-10 rated extensions -- according to the official \gls{joomla} extension directory -- as a representative set for further analysis. After choosing the set of extensions, we concentrated the scope of our security analysis on two main classes of injection bugs called \gls{xss} and \gls{sqli}.

\subsection{Detecting \glsentrylong{xss}}

\paragraph{Reflected \& Persistent \glsentrylong{xss}.}
We used two approaches to detect reflected and stored \gls{xss} as server-side vulnerabilities. First, we automatically identified all user inputs derived from HTTP GET and POST messages. We marked each input as \textit{secure} in case that special characters (e.g., the less than sign) are represented by HTML entities or a Unicode representation when it is returned within a response. Second, we scanned each extension for outputs that are generated by its views. We marked outputs as \textit{secure} when they were wrapped by escape functions such as \lstinline|htmlentities|, \lstinline|addslashes|, and \lstinline{$this->escape}. Every input or output that was \textit{not} marked as secure was inspected manually. We use an \lstinline|alert(1)| as a proof of concept to verify whether we have found an \gls{xss} issue.



\paragraph{\glsentrylong{domxss}.}
Modern applications like Google's GSuite allow to completely use complex apps within a web browser. To provide such a functionality, an enormous amount of JavaScript code is required. With the shift from desktop- to mobile-based applications, the capabilities of JavaScript code has also increased, and in this regard, the probability to carry out attacks as shown by Lekies et al.~\cite{Lekies:2013:MFL:2508859.2516703}.

In contrast to server-side vulnerabilities, \gls{domxss} allows a malicious party to attack a victim on the client-side without sending any suspicious HTTP requests to the server; this can be realized, for example, by using the hashtag within an \gls{xss} attack vector. To detect \gls{domxss} vulnerabilities, we are using the methodology provided by di Paola~\cite{domxssStefano}. By reusing the provided regular expressions as search patterns, we looked for sources (e.g., \lstinline|document.location|) and sinks (e.g., \lstinline|document.write|) within the \gls{dom} that could introduce \gls{domxss} vulnerabilities. After finding a suspicious sink or source, we move forward by generating different inputs and observing if, and how, they are modified. If we detect the reflection of an input value, we later also evaluated its exploitability.

\subsection{Detecting Metadata \glsentrylong{xss}} 
\label{sec:metadata-attacker}

Many extensions add the functionality of providing, displaying and editing photos and other multimedia files. Prominent examples in case of \gls{joomla} are Phoca Gallery, Ozio Gallery, and Sigplus.
Extensions can use metadata to display image captions or position information and it is thus of high importance to verify whether malicious and falsely trusted user input in media files can be used for \gls{xss} attacks. We could not identify any tool which fulfills our requirements of automatically putting \gls{xss} payloads within metadata fields.
Therefore, we have created a dockerized PHP-based open-source tool called \textit{\gls{metadata-attacker}}.  

Metadata is usually created by devices such as cameras which store information like GPS coordinates within media files (e.g., \lstinline|jpg|). As an example, one specification for saving metadata within images is provided by the \gls{iptc}. The specification allows the saving of data within self-descriptive properties like \textit{City} or \textit{Country}. This data is afterwards used by, for example, photographers to get more information about pictures (e.g., copyright data and the date of creation).

Instead of writing harmless data like IPTC properties within a media file, it is also possible to save arbitrary strings like \gls{xss} vectors in metadata fields. For this reason, we provide the penetration testing tool \textit{\gls{metadata-attacker}}, which allows to create an image (\lstinline|jpg|), audio (\lstinline|mp3|), or video (\lstinline|mp4|) file containing custom metadata or a set of \gls{xss} vectors to test any metadata interpreting application. Therefore, we have the possibility to test \gls{joomla}'s gallery extensions against possible \gls{xss} vulnerabilities when displaying unfiltered meta data (e.g., for detecting \gls{domxss}).

\gls{metadata-attacker} runs in a web browser and allows the selection of all, or just specific metadata fields that should be modified. To test for \gls{xss} vulnerabilities, our tool is able to automatically fill in -- by considering the length limit of each property -- a collection of over 200 well-established test vectors collected from the OWASP \gls{xss} Filter Evasion~\cite{owaspXSSCheatSheet} and HTML5 Security Cheat Sheets~\cite{html5secsheet}.

\begin{figure}[ht]
\centering
\includegraphics[width=\linewidth]{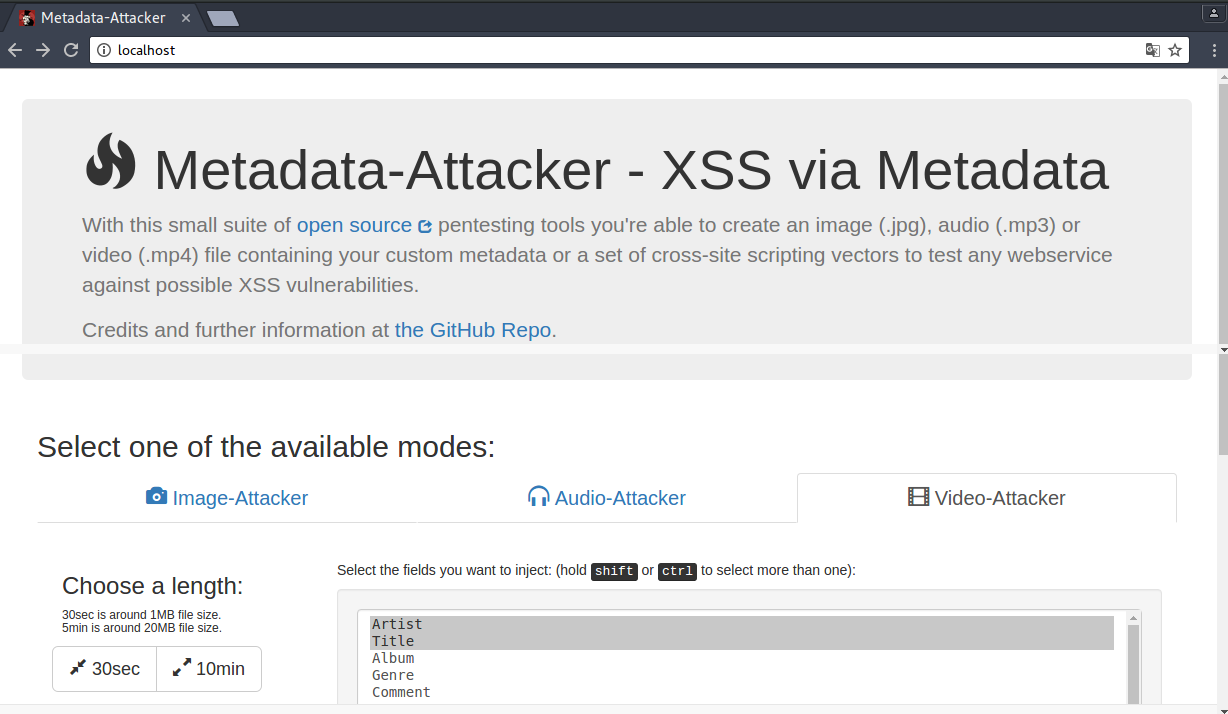}
\caption{Our open-source tool \glsentrytext{metadata-attacker} allows us to manipulate metadata of media files like images.}
\end{figure}

\subsection{Detecting SQL Injections}
Similarly to reflected and stored \gls{xss}, we used two approaches to detect \gls{sqli} vulnerabilities. First, we automatically identified all possible inputs derived by HTTP GET and POST messages. We marked each input as \textit{secure} if we cannot cause an \gls{sql} error or any change to the applications response; we have set the PHP flag \lstinline{error_reporting} to \lstinline|on| for debugging purposes.

Second, we scanned the extension for all \gls{sql} queries that are constructed with user input and potentially executed. We marked inputs as \textit{secure} if they were properly escaped upon usage with PHP functions such as \lstinline|$db->quote|, \lstinline|$db->quoteName|, or \lstinline{intval}. As in case of \gls{xss}, every input that was \textit{not} marked as secure was inspected manually.

\subsection{Security Issues \& Responsible Disclosure}
To identify the number of security issues, we only count a vulnerable parameter or variable once even if it might be reflected multiple times within the source code of a PHP file.

We reported all the discovered vulnerabilities to \gls{joomla}'s \gls{vel} team who forwarded our reports to the respective extension developers. Whenever it was explicitly requested, we worked closely with the developers to verify the correctness of their patches.
\section{Evaluation of Joomla Installations} 
\label{sec:noinstall}
\begin{table*}[ht!]
	\centering
	\begin{tabular}{l|l|c|c|c|c|c}
		Joomla! Extension & Identifying Path &  Installations & rXSS & sXSS & DOMXSS & SQLi \\ \hline \hline
		\rowcolor{Gray}	Akeeba Backup & \lstinline|com_akeeba| &  59.6\% & 3 & -- & -- & -- \\
		AcyMailing & \lstinline|com_acymailing|& 34.4\% & 5 & -- & -- &  1 \\
		\rowcolor{Gray}	Advanced Module Manager & \lstinline|com_advancedmodules| & 12.4\% & -- & 1 & -- & -- \\
		JEvents & \lstinline|com_jevents| &  11.9\% & 5 & 2 & -- & 1 \\
		\rowcolor{Gray} eXtplorer & \lstinline|com_extplorer| & 11.4\% & 6 & -- & -- & -- \\
		Phoca Gallery & \lstinline|com_phocagallery| & 4.7\% & 18 & -- & -- & -- \\
		\rowcolor{Gray}	Community Builder & \lstinline|com_comprofiler| & 3.6\% & 1 & -- & -- & -- \\
		Ark Editor & \lstinline|com_arkeditor| & 3.1\% & 2 & 5 & -- &  1 \\
		\rowcolor{Gray}	Ozio Gallery & \lstinline|com_oziogallery3| & 1.7\% & -- & 31 & -- & -- \\
		Sigplus & \lstinline|/media/sigplus| & 1.2\% & -- & 1 & 3 & --
	\end{tabular}
	\caption{Security analysis of the top-10 rated \gls{joomla} extensions in October 2018.
	The percentage of installations was calculated by considering all 7,797 detected \gls{joomla} installations.
	By looking of the numbers of \gls{xss} issues, rXSS describes reflected, sXSS stored, and DOMXSS DOM-based vulnerabilities.}
	\label{tab:xssSQLi}
\end{table*}

\subsection{Installed  \glsentrytext{joomla} Instances}
Among the \gls{alexa1m}, some hosts were offline and thus not reachable. Therefore, we successfully crawled 93.2\% of the \gls{alexa1m}.

During our scan for \gls{joomla} installations, we observed that some websites do not host \gls{joomla} within the standard root directory. Therefore, we also tried to discover the path of their installation. Whenever we could not detect an installation within the base directory, we sent a baseline request and followed all redirects. Afterwards, we extracted the path of the final destination and prepended it to our test paths.
Additionally, we detected that some websites behave very differently if the user agent doesn't belong to a browser.
Therefore, we used the user agent of Google Chrome 74 for our script.
To accommodate for network latency we set timeouts as high as ten seconds for a connection and 20 seconds for reading, while allowing up to ten retries in total.

All together, our crawler detected that over 2\% of the reachable Alexa websites use \gls{joomla}. Popular examples for websites with an active  \gls{joomla} installation are the National University of Singapore (\url{nus.edu.sg}), the Government of India (\url{uidai.gov.in}), and the Canadian Immigration (\url{canadavisa.com}).

\subsection{Installed \glsentrytext{joomla} Extensions}
In order to detect if an extension was installed within a default setup of \gls{joomla}, we verified the existence of specific static files within predictable installation paths (cf., \autoref{tab:xssSQLi}). For every extension except Sigplus, we scanned for an extension specific file (e.g., \lstinline|manifest.xml|) within the path \lstinline|/administrator/components/[identifier]|; the identifier was replaced with the component's name listed in \autoref{tab:xssSQLi}. Every file contains strings with content -- such as the extension's name -- with which we were able to clearly fingerprint each extension.

Sigplus is not a component and therefore, we had to verify the availability of the JavaScript file \lstinline{initialization.js} within the path \lstinline{/media/sigplus/js/}. Analogously to our fingerprinting approach for components, the detected JavaScript file also contains unique strings to identify the extension.

In total, we detected 7,797 \gls{joomla} installations with extensions. As listed in \autoref{tab:xssSQLi}, the most popular extension Akeeba Backup was included in 4,646 detected installations; it is thus available in nearly 59.6\% of our scanned \gls{joomla} installations. AcyMailing is the second most popular extension with a market share of 34.4\% within our tested installations. Among the top-10 rated extensions, Ozio Gallery and Sigplus are the most unpopular extensions with an availability of less than 2\%.

Our results also show that the top-10 rated extensions are not equal to the top-10 of the most popular extensions. \gls{joomla} does not collect data about the number of installation and therefore, there is no official list about the most popular extensions.

\section{Security Evaluation} 

As displayed in \autoref{tab:xssSQLi}, our evaluation revealed that we can identify \gls{xss} vulnerabilities in every single extension. In most of the cases, our methodology allowed us to find reflected \gls{xss} issues; 40 out of 83 vulnerabilities were vulnerabilities within the stored \gls{xss} area.

By using our tool Metadata-Attacker, \gls{domxss} vulnerabilities could be detected in the photo and multimedia gallery extension Sigplus. 
The vulnerabilities were caused by the unfiltered reflection of \gls{iptc} metadata provided by images (e.g., image headline data that we have replaced with JavaScript code).


We identified that the root cause of this vulnerability was not just in Sigplus; the JavaScript libraries \textit{boxplus} and \textit{slideplus}, which are shipped with Sigplus, are also affected.

Over 44\% of our discovered \gls{xss} vulnerabilities can be used to attack a victim which does not have any special privileges. Our detected \gls{sqli} vulnerabilities can be only exploited with an account which has administrative privileges. For example, ArkEditor's \gls{sqli} vulnerability can be only used in the event that the attacker can edit plugins; AcyMailing's \gls{sqli} vulnerability requires access to the statistics page of the AcyMailing management interface.

\subsection{\textit{Case \#1:} Reflected \glsentrytext{xss} in JEvents}
This vulnerability within JEvents was detected with our second approach to discover reflected and stored \gls{xss}. We first discovered the following potentially vulnerable code 
\lstinline|<input type="text" [...] value="<?php echo $this->keyword;?>" />| within the component file  \lstinline|form_body.php|.
Due to the reason that the parameter \lstinline|$this->keyword| is echoed directly without any escaping mechanism, we further investigated its origin.
Our investigation led us to the components controller file (\lstinline|search.php|) in which the code displayed in \autoref{lst:XSS_1} is used to process the input parameter \lstinline|keyword|.

At the beginning of \autoref{lst:XSS_1}, the request parameter \lstinline|keyword| is saved in the variable \lstinline{$keyword}.
Thereafter, if the input is longer than the predefined maximum length of 20 characters, it is truncated to a length of 20 characters.
However, instead of using the already truncated value, JEvents uses the original input value combined with a protection against \gls{sqli} vulnerabilities (\lstinline|db-escape|, cf. line 6).

The originally truncated value is therefore overwritten.
Most importantly, the escaping is applied for a database context instead for an \gls{html} attribute context.
Finally, the value that is escaped in order to avoid \gls{sqli} is assigned to the view without any truncation.


\begin{sloppypar}
As an exploitation example, a malicious URI including \lstinline[keepspaces]|/index.php/searchjevents?keyword=XSS" onfocus=alert(1) autofocus=| could be used to trigger an \gls{xss} payload; we break out of the attribute context with a double-quote sign and automatically execute arbitrary JavaScript code by triggering the event handler \lstinline|onfocus|. Based on the privileges of the attacked user, it is possible to attack the complete \gls{joomla} installation as \gls{joomla}'s frontend and backend reside on the same origin (e.g., to get administration privileges).
\end{sloppypar}


\begin{lstlisting}[language=php, basicstyle=\ttfamily\footnotesize, label={lst:XSS_1}, caption={Vulnerable processing of the parameter \textit{keyword}. The code is protecting the user against \gls{sqli} attacks and not against attacks in the HTML context.}]
$keyword = $jinput->getString('keyword', '');
$upper_limit = 20;
if (JString::strlen($keyword) > $upper_limit) {
  $keyword	= JString::substr($keyword, 0, $upper_limit - 1);
}
$keyword = $db->escape($jinput->getString('keyword', ''));
[...]
$this->view->assign("keyword",$keyword);
\end{lstlisting}

\subsection{\textit{Case \#2:} Persistent \glsentrytext{xss} in Ark Editor}

The first approach of our methodology allowed us to identify a stored \gls{xss} vulnerability in Ark Editor by testing all inputs that we could observe. In general, Ark Editor provides the functionality of modifying content directly on a \gls{joomla} page by allowing front-end inline editing. Whenever an article is edited with Ark Editor's editing tool, an \gls{ajax} request is sent to the back-end. This HTTP request includes the parameter \lstinline|data[articletext]|, which contains the Base64 encoded content of the whole article.

The intended behaviour of Ark Editor is to let the user of the editor modify content, which is always encoded with HTML entities; therefore, greater and lesser than signs that are, for example, required to inject an HTML element like \lstinline|<script>|, and will not be parsed by the browser. However, a user still has the possibility to click on the editor bar and thus inject elements like images. This occurs because the editor places an \gls{html} \lstinline|<img>| tag in the background and is displayed as an image in the front-end editor.

Our observation shows that the Base64 encoded article, including unescaped HTML code, is transmitted within the parameter \lstinline|data[articletext]|. We have, therefore, replaced the Base64 encoded article content with a Base64 encoded \lstinline|<script>alert(1)</script>| string. As a result, an alert-window is always displayed when the article is shown to a user.

Usually the default text filter is applied in \gls{joomla}, which does not allow any user besides the super user to include unsafe \gls{html} in an article. However, Ark Editor's security only relied on the client-side escaping of special characters within the front-end editor itself. Moreover, Ark Editor did not implement any \gls{csrf} protections in this case, leading to a by Ark Editor self-scored \textit{high risk} \gls{xss} vulnerability~\cite{arkXSSBlog}.

\subsection{\textit{Case \#3:} \glsentrytext{domxss} in Sigplus}

With the help of our tool \gls{metadata-attacker}, we ware able to detect \gls{domxss} vulnerabilities within the photo and multimedia gallery Sigplus. 

We used \gls{metadata-attacker} to create a sample image containing an \gls{xss} payload in every metadata field. Afterwards, we uploaded the image to the gallery and, thereafter, opened it in a web browser. With this approach, we are able to trigger four \gls{xss} vulnerabilities leading to JavaScript-based alert windows; three of them were \gls{domxss} and one of them stored \gls{xss} issues. One particular case for a \gls{domxss} issue was caused by the combination of Sigplus and, on of the JavaScript libraries it is shipped with, Slideplus.

Initially, Sigplus's backend code extracted some metadata fields from the image (e.g., the image headline) and stored them in a \lstinline|noscript| element; some metadata fields were encoded with PHP's \lstinline|htmlentities| function.
Afterwards, the client-side initialization script of Sigplus used JavaScript's \lstinline|innerHTML| to set the \gls{html} content of a \lstinline|div| element.
From there, Slideplus extracted the image headline, which at this point was still encoded with \lstinline|htmlentities|, and unescaped it.
This unescaped value was then used in conjunction with \lstinline|innerHTML|, which lead to the \gls{xss} vulnerability.
Hence an image with the IPTC field number 105 (image headline) set to \lstinline|<img src=x onerror="alert(1)">| could be used to create an alert window; in our example, an image with a not existing resource triggers the \lstinline|error| event and executes JavaScript code as a proof of concept.

\subsection{\textit{Case \#4:}  \glsentrytext{sqli} in JEvents}
While following our second approach to discover \gls{sqli}, we discovered the following code in the JEvents library file  \lstinline{saveIcalEvent.php} (cf., \autoref{lst:SQLI1}).

We first detected the \gls{sql} query, which uses the apparently unsanitized value \lstinline{$ics_id}.
By backtracing the input array, we discovered that it is only filtered for the HTML context but not escaped for the usage in an \gls{sql} query; before this filtering step the array is directly read from the request.
Our analysis has, therefore, shown that a request which contains valid post data to create or edit an event can be used to execute an \gls{sqli} attack.


\begin{lstlisting}[language=php,basicstyle=\ttfamily\footnotesize, label={lst:SQLI1},caption={By saving an event, an attacker can use POST data manipulation to trigger an \gls{sqli}.}]
$ics_id=ArrayHelper::getValue($array, "ics_id",0);
[...]
$vevent->catid = 
  ArrayHelper::getValue($array,  "catid",0);
[...]
if ((is_string($vevent->catid) && $vevent->catid<=0) || (is_array($vevent->catid) && count($vevent->catid)==0)){ 
  $query = "SELECT catid FROM #__jevents_icsfile WHERE ics_id=$ics_id";
  $db->setQuery( $query);
  $vevent->catid = $db->loadResult();
\end{lstlisting}

As an exploitation example, we can use the query \lstinline|1 union select extractvalue(rand(),concat(0x3a,(select "SQLi")))| as a value for the parameter \lstinline|ics_id| to do an \gls{sqli} attack; we add an additional select and extract arbitrary data from the current database.
As \gls{joomla} uses only a single database for all operations our injection grants read access to the complete installation.
\section{Discussion and Mitigations}

\subsection{Impact on other CMS Systems}
Due to the high number of bugs in \gls{joomla} extensions, we also looked for extensions that are available for other \gls{cms}s, like \gls{drupal} and \gls{wp}. Our analysis has shown that two extensions from our testbed are also available for other systems.

First, eXtplorer is comparable to a standalone web-based file management component and, thus, also available as a \gls{wp} and \gls{drupal} extension. Due to exactly the same source code, we have the same \gls{xss} vulnerabilities on every large \gls{cms} system which allows developers to include this component as an extension. Second, Akeeba Backup is also available for \gls{wp}. Although the source code is modified to be compatible with \gls{wp}, we could identify that it has the same structure and, therefore, the same vulnerabilities.

In total, we could derive 17 new \gls{xss} vulnerabilities by looking on eXtplorer and Akeeba Backup.

\subsection{Vulnerability Root Causes} 
Although \gls{joomla} provides a list of security guidelines with the official documentation, we detected three different root causes for our security issues within the \gls{xss} and \gls{sqli} area.

First, most of the extensions are vulnerable because of missing filtering mechanisms. We could, therefore, assume that developers might be unaware of our detected security vulnerabilities, or that they have simply forgotten to implement a security mechanism. Moreover, \gls{joomla} does not provide any official filtering mechanism against \gls{domxss} attacks. 

Second, some extensions secure the user's input and output by using self-written sanitizers that can be sometimes bypassed. For example, Akeeba Backup explicitly escaped special characters like the single quote within an inline \lstinline|script| element; as a bypass to break out of the JavaScript context, we have injected the closing \lstinline|script| element with greater and less than signs that were not escaped.

Third, some developers (cf., \autoref{lst:XSS_1} and \ref{lst:SQLI1} for JEvents) are aware of security vulnerabilities but they are protecting their application by misusing security features that are provided by \gls{joomla}. For example, JEvents protected the user against \gls{sqli} although the protecting context was actually \gls{xss}; therefore, some characters which could be used to break out of the HTML attribute context were not filtered by \gls{joomla}.

\subsection{Code Analysis Tools} 
An approach to secure \gls{joomla}'s extension directory could be to establish a manual review process to analyze an extension before it is published in the extension store. However, \gls{joomla} is an community-based open-source project with limited manpower and thus, it might be more efficient if this review process is done automatically on the extension's code.

Since June 2018, \gls{joomla}'s code base is continuously scanned by the code analysis tool RIPS to detect security vulnerabilities~\cite{joomlaNewsRips}. In contrast to the core systems, extensions are currently not scanned by such code analysis tools before they are published. To achieve the same level of security, extensions should be scanned by similar tools as well.

It is important to note that code analysis tools only detect the lower bound of the existing vulnerabilities. For example, during our research study we also detected two stored \gls{xss} vulnerabilities within the core itself, even though RIPS and NAVEX~\cite{217650} were previously used as code analysis tools, and having a focus on \gls{xss} (CVE-2019-6262, CVE-2019-9712).

Nonetheless, code analysis tools might help to reduce the number of vulnerabilities. Due to our bug reports, we learned that in some cases, the reported bugs were only partially fixed. For example, the popular backup extension Akeeba Backup only fixed two out of three \gls{xss} bugs correctly on the first attempt; one bug was only fixed for HTTP GET requests while attacks via HTTP POST were still possible. 

\subsection{Sandboxing via CSP \& Trusted Types} Browsers like \gls{gc} and \gls{ie} introduced \gls{xss} prevention tools called \gls{xss} filter (\gls{ie} 8) and \gls{xss} Auditor (\gls{gc} 4)~\cite{Bates2010RegularEC}. The aim of these tools is to detect \gls{xss} attacks within the browser itself. In case of \gls{gc}, \gls{xss} auditor attempts to find reflections during the HTML parsing phase and therefore it blocks the execution of potential malicious code like JavaScript. Researchers have shown that these prevention mechanisms could be used as a tool to do cross-site information leak attacks and therefore browsers vendors have decided to remove such filters (e.g., in \gls{gc} 78)~\cite{xssAuditorRemove, xssAuditorPortSwigger}.  Due to the reason that modern browsers like \gls{gc} are no more able to detect even simple reflected \gls{xss} attacks, there should be a special focus on sandboxing mechanisms like \gls{csp} and Trusted Types.

\gls{joomla} allows the activation of the \gls{csp} to primarily mitigate content injection vulnerabilities such as \gls{xss}. Although the core system can be protected with \gls{csp}, \gls{joomla} extensions do not provide a fine granular \gls{csp} ruleset. A major challenge would be to automatically verify -- when the extension is uploaded to the extension store -- \gls{csp} rules regarding their securenes. Weichselbaum et al.~\cite{45542} showed that 99.34\% of the scanned hosts with a CSP do not offer a benefit against \gls{xss}; inter alia, due to the usage of \lstinline|unsafe|-directives. As a possible verification tool to face this problem, \gls{joomla}'s extensions store could make use of Google's \gls{csp} Evaluator.\footnote{\url{https://csp-evaluator.withgoogle.com/}}


Trusted Types~\cite{TTDraft, TTGithub} are a new and experimental feature which has been proposed to tackle \gls{domxss} vulnerabilities. This feature can be enabled by using the \gls{csp}.
The main idea is to create special JavaScript object types that have to be assigned to return potential \gls{domxss} sinks; otherwise the browser would throw a \lstinline|TypeError| in the event that an untrusted value is returned. The only way to create those new types is by implementing policies which can be defined by using the Trusted Types browser API. While this does not automatically prevent \gls{domxss}, it enforces a check of sinks (e.g., \lstinline|location.hash|) and makes code audits easier; only the newly defined Trusted Types policies have to be reviewed for security issues. A major drawback is that existing legacy code needs to be reworked to apply specific policies for each use case.
However, in order to allow a continuous migration of legacy code, a default policy can be defined which is applied if an untrusted string would be assigned to an injection sink. Therefore, Trusted Types could be used within a \gls{cms} like \gls{joomla} while the core, and also the extensions, could be migrated continuously.

\section{Related Work}

\subsection{CMS Security}
In \citeyear{Trunde:2015:WSA:2837185.2837195}, \citet{Trunde:2015:WSA:2837185.2837195} created an analysis based on publicly available exploits within Wordpress by nearly taking 200 vulnerabilities into account. As a contribution, they outlined that a combination of manual and static analysis might be the best solution to discover similar vulnerabilities; we used the same approach to discover \gls{xss} and \gls{sqli} vulnerabilities.
Extensions were not within their scope.

In \citeyear{VanAcker:2017:MLW:3019612.3019798}, \citet{VanAcker:2017:MLW:3019612.3019798} evaluated the security of login forms by looking on popular web frameworks and \gls{cms} (e.g., \gls{joomla}, \gls{drupal} and \gls{wp}). Their evaluation of the Alexa top-100 thousand showed that the implemented security measures were lacking in numerous cases because \gls{hsts}, \gls{hpkp}, \gls{sri}, and various \gls{csp} directives, such as \gls{uir} and \gls{bamc}, were not implemented.
Injection attacks such as \gls{xss} and \gls{sqli} were not considered.

\subsection{\glsentrytext{cms} Extensions}
In \citeyear{6983376}, \citet{6983376} analyzed the security of 35 \gls{wordpress} extensions using the static code analysis tools phpSAFE and RIPS.
They found more than 350 \gls{xss} and \gls{sqli} vulnerabilities and thereby showed that plugins in \gls{wordpress} pose a security risk.
They argued that prior to its release the security of a plugin should be verified through static code analysis by the developers and the core application providers.
They neither considered other \glspl{cms}s, nor detected cross-\gls{cms} bugs.

In \citeyear{7528327}, \citet{7528327} argued that while \glspl{cms} provide a lot of flexibility, they also result in more attacks and additional vulnerabilities through extensions.
In addition to common security best practices, they proposed a centralized reporting tool which forwards reports of security vulnerabilities to the correct party to reduce the amount of vulnerabilities introduced through the usage of a \gls{cms}.

In \citeyear{metadata_xss_joomla}, \citet{metadata_xss_joomla} detected a persistent \gls{xss} in a \gls{joomla} extension using image metadata.
He used the \texttt{exif} commandline tool to include the \gls{xss} payload within the \texttt{Caption-Abstract} metadata.

In \citeyear{8744951}, \citet{8744951} evaluated the security of the most common backup plugins for \gls{wordpress}.
They showed that 12 out of 21 investigated extensions leak sensitive data, for example, by making backups accessible to everyone.
They concluded that most backup plugins fail to use proper randomness and strong cryptographic methods, and therefore end up posing a security risk.

In the same year, \citet{Ruohonen:2019:DVS:3319008.3319029} pointed out that \gls{wordpress} as the most popular \gls{cms} \enquote{has had a particularly bad track record in terms of security}, but in the recent years the security vulnerabilities were more likely to be found in its plugins.
He studied the correlation between the popularity of a plugin and the amount of reported and disclosed vulnerabilities, and through an empirical study concluded that those two properties are indeed correlated.

\subsection{Systematic Scanning for Web Vulnerabilities}
There is a large body of research considering \gls{xss} and \gls{sqli}. In the following paragraphs, we highlight the prevalence of these vulnerabilities by performing systematic scans.

In \citeyear{Lekies:2013:MFL:2508859.2516703}, \citet{Lekies:2013:MFL:2508859.2516703} evaluated the prevalence of DOM-based XSS attacks in Alexa top 5,000 websites.  In their study, they identified over 6,000 unique vulnerabilities distributed over 480 domains. Nearly 10\% of the analyzed domains had at least one \gls{domxss} issue.

In \citeyear{Parameshwaran:2015:DRT:2786805.2803191}, \citet{Parameshwaran:2015:DRT:2786805.2803191} introduced the tool \texttt{DEXTERJS} which uses instrumentation to perform taint analysis on JavaScript code supplied by a target website.
They crawled the Alexa top 1,000 and, using their tools, found 820 distinct zero-day \gls{domxss} vulnerabilities.
The systematic analysis of \gls{domxss} has been in the scope of many additional researchers (e.g.,~\citet{stock2014precise,DBLP:conf/ndss/SteffensRJS19,stock2017web}).

In the same year, \citet{7412085} evaluated the ability of three black-box scanners to find \gls{sqli} and \gls{xss}.
They concluded that the scanners still fail to find all vulnerabilities in their testbed.

In comparison to all these works, we identified \gls{xss} and \gls{sqli} vulnerabilities directly in the analyzed \gls{joomla} extensions. Afterwards, we estimated their impact by scanning for the frequency of the vulnerable extensions in all \gls{alexa1m} websites. Our approach allowed us to perform large-scale scans on more websites and responsibly disclose our vulnerabilities directly to the extension developers, whose fixes had a direct impact on all the analyzed websites.

\section{Concluding Remarks}

We presented a security analysis of the top-10 rated \gls{joomla} extensions. We showed the generalizability of detecting cross-platform vulnerabilities on \gls{drupal} and \gls{wp}. Our methodology allowed us to identify \gls{xss} vulnerabilities in each of the analyzed extensions, and \gls{sqli} vulnerabilities in 30\% of the analyzed extensions. Due to a missing isolation between the core system and the installed extensions, these vulnerabilities had a direct impact on the security of the whole system. 
In total, the detected vulnerabilities affected over 40\% of the \gls{alexa1m} websites using \gls{joomla}.  


The direct security impact of core extensions shows that highly-ranked extensions must receive as much attention as the \gls{joomla} core system. Every extension should be critically reviewed before being published in the official directory. This should be done by dedicated security audits or carefully implemented automated security tools.

From a scientific point of view, it is also interesting to study the missing separation between the \gls{joomla} core and extension system. New research directions could study the possibility regarding the separation of both technologies, and develop multi-stage concepts of separating accessible data from the extensions and the core system. 

Overall, our findings prove that these directions should not only be followed by \gls{joomla}, but additionally, similar techniques and research directions should be studied in other \glspl{cms}, such as \gls{drupal} and \gls{wp}.

\bibliographystyle{plainnat}
\bibliography{bib/refs}

\end{document}